\documentclass[twocolumn]{aastex631}
\usepackage{hyperref}
\usepackage{color}
\usepackage{amssymb}
\usepackage{amsmath}
\usepackage[ruled,linesnumbered]{algorithm2e}
\usepackage{footmisc}
\usepackage{ulem}

\shorttitle{Sedona IBT}
\shortauthors{Chen et al.}
\begin{document}

\title{An Integral-Based Technique (IBT) to Accelerate the Monte-Carlo Radiative Transfer Computation for Supernovae}
\correspondingauthor{Lifan Wang}
\email{lifan@tamu.edu}

\author[0000-0003-3021-4897]{Xingzhuo Chen}
\affiliation{George P. and Cynthia Woods Mitchell Institute for Fundamental Physics \& Astronomy, \\
Texas A. \& M. University, Department of Physics and Astronomy, 4242 TAMU, College Station, TX 77843, USA}

\author{Lifan Wang}
\affiliation{George P. and Cynthia Woods Mitchell Institute for Fundamental Physics \& Astronomy, \\
Texas A. \& M. University, Department of Physics and Astronomy, 4242 TAMU, College Station, TX 77843, USA}

%\author{Yi Yang}
%\affiliation{Department of Astronomy, University of California, Berkeley, CA 94720-3411, USA}

%Gesa: I am adding a collaborator in. Dr. Yi Yang is an expert in supernova spectropolarimetry observation and data reduction. He could provide some observational data so that I could use the current toy model to fit an observation. 

\author{Daniel Kasen}
\affiliation{Astronomy Department and Theoretical Astrophysics Center, University of California, Berkeley, Berkeley, CA 94720, USA}

\begin{abstract}

We present an integral-based technique (IBT) algorithm to accelerate supernova (SN) radiative transfer calculations. 
The algorithm utilizes ``integral packets'', which are calculated by the path integral of the Monte-Carlo energy packets, to synthesize the observed spectropolarimetric signal at a given viewing direction in a 3-D time-dependent radiative transfer program. 
Compared to the event-based technique (EBT) proposed by \cite{Bulla2015VPack}, our algorithm significantly reduces the computation time and increases the Monte-Carlo signal-to-noise ratio. 
Using a 1-D spherical symmetric type Ia supernova (SN Ia) ejecta model DDC10 and its derived 3-D model, the IBT algorithm has successfully passed the verification of: (1) spherical symmetry; (2) mirror symmetry; (3) cross comparison on a 3-D SN model with direct-counting technique (DCT) and EBT. 
%DK: Not clear what you mean by passing the verification of "spherical symmetry". Do you mean that you get the correct solution in both 1D and 2D / 3D simulations?  
%Gesa: yes, that is what I mean. 
Notably, with our algorithm implemented in the 3-D Monte-Carlo radiative transfer code SEDONA, the computation time is faster than EBT by a factor of $10-30$, and the signal-to-noise (S/N) ratio is better by a factor of $5-10$, with the same number of Monte-Carlo quanta. 
%LW: Do you mean the coomputation time is faster by a factor of $10-30$ and the SNR is better by a factor of $5-10$? Why don't you just say how many times faster for the same SNR?
%DK: May be better worded to say that we can increase the S/N by a factor of 5-10 with only a 30% increase in computational time.
%Gesa: I have rewritten the text to make a more clear comparison of the computation time and the S/N between IBT and EBT. 
%Specifically, we use a 3-D ejecta model derived from the 1-D spherical symmetric model DDC10 to verify that our algorithm could generate same or reversed spectropolarimetry signal on mirror symmetric viewing directions. 
%We also made cross comparison between the spectropolarimetry results from EBT and IBT on the 3-D ejecta model. 
%To compare the Monte-Carlo signal-to-noise ratio between the algorithms, we use the 1-D DDC10 model to verify if the algorithms could reproduce the theoretical zero-polarization result, and we demonstrate that the signal-to-noise ratio from IBT is higher than EBT by a factor of $5-10$. 

\end{abstract}


\section{Introduction}\label{sec:intro}

Type Ia supernovae (SNe Ia) have been widely applied in cosmological studies (e.g. \cite{Abbott2019Cosmology,Brout2022Pantheon}) based on the empirical relation between the light curve properties and the luminosities (e.g. \cite{Phillips1993Relation}). 
However, the explosion mechanism of SNe Ia is still a mystery;  studies of the ejecta structure  rely on simulations that include hydrodynamics, nucleosynthesis, and radiative transfer. 

Radiative transfer simulations of supernovae calculate the specific intensity and plasma excitation states to obtain spectra that can be compared to optical spectroscopic and photometric observations. 
Many radiative transfer codes assume that SNe Ia are spherical symmetric and reasonable agreement have been found among the simulation results of different codes \citep{Blondin2022Standart}. 
%For example, CMFGEN \citep{Hillier1998CMFGEN,Hillier2012CMFGEN} is an 1-D time-dependent radiative transfer program that directly solves the radiative transfer equation, and a further probe of the solution in 3-D is provided in the PHOENIX program \citep{Hauschildt2006Phoenix}. 
%DK: I would not mention just the CMFGEN and PHOENIX codes, as there are many SN radiation transport codes. So to mention just a few appears to neglect many other people's work. I think it is better to not cite any, and maybe just mention that many codes exist and that relative agreement is found between them, citing the Blondin paper which collects many codes
%DK: I wouldn't say that MC methods "accelerate" the computation. ACtually, MC methods can be rather slow. But MC do provide a flexible way to treat complicated physical processes within the transport code.

%DK: I would move the statement about the agreement of different transport codes to where I mentioned in the above paragraph.

%Gesa: The sentences are rewritten, only the Blondin's comparison paper is cited here. 
However, imaging of nearby SN Ia remnants (e.g. \cite{Ferrazzoli2023Tycho}) and hydrodynamic simulations (e.g. \cite{Gamezo2004DDT,Kromer2013Hesma}) suggest that SNe Ia have an asymmetric 3-D structure. 
Given the distance to most SNe Ia, this 3-D structure cannot be resolved, but can be probed by spectropolarimetric observations \citep[e.g.,][]{Wang_2003,Cikota2019PolObs,Yang_2022}. 
The polarization from SN Ia is primarily due to Thomson scattering of photons, and a non-zero linear polarization signal can arise in asymmetric SN ejecta due to the incomplete cancellation of polarization from photons scattered from different geometrical structures \citep{Wang2008SpecPol}. 

Monte-Carlo methods, which use energy packets to emulate the propagation of photons in the SN plasma, provide a flexible way to treat complicated physical processes \citep[][]{Lucy2002macroatom,Lucy2003macroatom2} and have been applied in 3-D time-dependent radiative transfer codes such as SEDONA \citep{Kasen2006Sedona}, and ARTIS \citep[][]{Kromer2009ARTIS}. 
3-D Monte-Carlo radiative transfer (MCRT) simulations can be computationally expensive, as many photon packets must be propagated to reduce the statistical noise in the synthesized observables. 
This especially poses a challenge for modeling SN polarization, as the expected signals are typically only at the 1\% level or less. 
To reduce the noise in SN MCRT simulations, \cite{Bulla2015VPack} introduced an ``event-based technique'' (EBT) for spectropolarimetry synthesis, which has been applied to simulations of SNe Ia  \citep{Bulla2016merger,Bulla2016dd}, superluminous SNe \citep{Inserra2016SLSN}, and kilonovae \citep{Bulla2021kilonova}. 
% DK: I think the Bulla method is what is referred to as EBT. If so, you should mention that here. 
%Gesa: I have corrected it. 

In this paper, we introduce a new integral-based technique (IBT), which is implemented in the MCRT code SEDONA, 
%DK: You have used the MCRT abbreviation above, but only defined it here. Be sure to define it before you use it. 
%Gesa: I have checked the definition of MCRT and make sure the definition is made the first time to use this phrase. 
to efficiently construct observable spectra from a 3-D MCRT computation. 
We find that IBT is $10-30$ times faster and $5-10$ times less Monte-Carlo noise than EBT, with the same number of Monte-Carlo quanta. 
%DK2: I was a little unclear is it both 10-30 times faster *and* less noise, or one or the other? 
%Gesa2: It is both. 
%LW: If it is both, rewrite it for the same noise level.
Moreover, IBT can be applied to the ultraviolet and infrared wavelength, and early phase of SNe, which was not calculated with EBT \citep{Bulla2015VPack}. 
% DK2: I am not sure how the application to early times or broader wavelengths is related to memory concerns. You may want to clarify. 
%Gesa2: the memory requirement of EBT is related to the number of OP scattering events per time step. There are more OP scattering events per time step at early phase of SNe and in the UV wavelength, due to the high central density and the high temperature. Bulla et al. 2015. only start the EBT calculation at 3500 Angstrom and at 10 days probably due to the limited memory space. It may not be explained with one sentence in the introduction section, there are more discussions in Section 2.2 and Section . 
Section \ref{sec:SpecExtract} reviews the direct-counting technique (DCT) and event-based technique (EBT) that have been used in previous MCRT codes. 
%which are the two spectropolarimetry synthesis algorithms reported in \cite{Kasen2006Sedona} and \cite{Bulla2015VPack}. 
Section \ref{sec:ibt} presents the IBT method. 
Section \ref{sec:validation} uses several toy models to verify the spectropolarimetry results from IBT. 
Section \ref{sec:compperf} compares the Monte-Carlo simulation error and computation time of the three algorithms: DCT, EBT, and IBT. 
Section \ref{sec:conclusion} summarizes our main conclusions. 
%DK: You may want to briefly mention the main point here, that your IBT method provides better results than DCT and EBT. 
%Gesa: I mentioned the computation time and S/N of the IBT and EBT here. 

\section{Polarized Spectrum Extraction}\label{sec:SpecExtract}
%DK: I'm not sure the title "Spectrum Extraction" is too informative. A title like "Polarized Monte Carlo Transport" or such may be be better 
%Gesa: The theme of this section is describing the methods to obtain observable spectra from MCRT, which is appropriately described by 'spectrum extraction'. If the title is 'polarized monte carlo transport', then the theme of the section will more emphasize on how the OPs undergo thomson scattering or even zeeman effect to have polarization signal. I have changed the title to 'polarized spectrum extraction'. 

In the SEDONA code, 
%the gamma-ray energy packets are created from the decay of radioactive isotopes, then converted to optical energy packets through Compton scattering and photoelectric absorption processes. 
%DK: I don't think you need to mention gamma-ray packets here, as they are not so relevant to the polarization modeling. Also, in Sedona you don't always need to produce OPs from gamma-rays; there are other ways, such as boundary emiters or point sources. 
%Gesa: Indeed, there is no need to mention gamma-ray packets. Will need to check the paper, make sure gamma-ray packets or GP is not mentioned. 
an optical photon packet (OP) represents a collection of photons with the same frequency, spatial coordinate, and propagation direction. 
The OPs undergo interaction events due to Thomson scattering, bound-bound transitions, bound-free transitions, and free-free transitions which can change their frequency and direction. 
%Following \cite{Lucy2002macroatom}, the OPs are not destroyed or split during the as-mentioned scattering and transition events. 
%Moreover, the energy of the OPs in the co-moving frame (where the SN plasma is at rest) are not changed before and after the interactions to guarantee energy conservation. 
%The trajectories of the OPs in a volume cell and in a time step is used used to calculate the mean intensity in the co-moving frame $\Bar{J_{\nu}}$, with the following equation: 

%\begin{equation}
%    \Bar{J_{\nu}}(\Vec{x},t_{exp})=\frac{\alpha_{abs}(\Bar{\nu})}{4\pi \Delta V \Delta t \Delta \Bar{\nu}} \sum_{i,j} \ \Bar{E_{i}}d_{i,j},
%\end{equation}

%where $\alpha_{abs}(\Bar{\nu})$ is the absorption opacity at the co-moving frame frequency $\Bar{\nu}$, $\Delta V$ is the volume of the cell, $\Delta t$ is the time step, $\Delta \Bar{\nu}$ is the width of the frequency grid, $\Bar{E_{i}}$ is the co-movining frame energy of the i-th OP, $d_{i,j}$ is the length of the j-th line segment of the polyline trajectory. 
%The summation only account for the trajectory line segments that satisfy the co-moving frequency of the OP falling in $[\Bar{\nu}-\Delta\Bar{\nu}/2,\Bar{\nu}+\Delta\Bar{\nu}/2)$. 

The polarization information of an OP is stored as a dimensionless Stokes vector 

\begin{equation}
    \boldsymbol{s}_{OP}=\begin{pmatrix}1 \\ q \\ u \\ v\end{pmatrix}\ , 
\end{equation}
and the Stokes vector can be constructed with $\boldsymbol{s}_{OP}$ and the energy of the OP $E_{OP}$. 
%and the Stokes vector can be expressed as: 
%DK: I think technically the Stokes vector would have units of intensity, where as your S here has units of energy. I don't think you need to define S here. It may be find to define the dimensionless one above, and the energy of the packet. 
%Gesa: The Stokes vectors with energy units are used in the appendix to explain the rayleigh scattering computation process in IBT. 
%\begin{equation}
%    \boldsymbol{S}_{OP}=\begin{pmatrix}I \\ Q \\ U \\ V\end{pmatrix}\ = E_{OP}\boldsymbol{s}_{OP}, 
%\end{equation}
%where $E_{OP}$ is the energy of the OP. 
%DK2: The Stokes vector I think traditionally has units of intensity, whereas this looks to have units of energy. An equation with the correct units is Eq 3. So you might not want to include this equation (2), but rather just say that the energy of the packet is E_op which can be used to construct a STokes vector intensity. 
%Gesa2: I have moved this equation into appendix. 

Typically, OPs are generated with zero initial polarization as expected for thermal emission. 
During the propagation of an OP, the linear polarization state changes in a Thomson scattering event according to the Rayleigh scattering phase matrix (Section 1.17, equation 217 of \citep{Chandra1960RT}). 
In bound-bound, bound-free, and free-free interactions, the OPs are typically assumed to be fully depolarized. 
The current version of SEDONA does not include magnetic fields, and the $V$ parameter in the Stokes vector, which represents the circular polarization, is set to zero. 

\subsection{Direct Counting Technique}

A straightforward method of spectral synthesis in MCRT is the direct counting technique (DCT), which sums the energy of OPs that escape the simulation domain within a certain frequency, time, and viewing angle bin. 
The observed flux is then calculated as 

\begin{equation}
    \begin{pmatrix} \mathcal{I} \\ \mathcal{Q} \\ \mathcal{U} \\ \mathcal{V} \end{pmatrix} = \frac{1}{4\pi r^2 \Delta t \Delta \nu \Delta \Omega} \sum_i E_{OP,i} \boldsymbol{s}_{OP,i}
\end{equation}
where $\boldsymbol{s}_{OP,i}$ is dimensionless Stokes vector of the $i$-th OP, $E_{OP,i}$ is the lab frame energy of the $i$-th OP, $\Delta t$ is the size of the arrival time bin, $\Delta \nu$ is the size of the frequency bin, $r$ is the distance to the SN center, $\Delta \Omega$ is the viewing direction bin size. 
The summation is over all the OPs in the arrival time interval $[t-\Delta t /2, t+\Delta t /2)$, the frequency interval $[\nu - \Delta \nu /2, \nu + \Delta \nu /2)$, and the viewing direction interval $[\varphi -\Delta \varphi/2, \varphi +\Delta \varphi/2)$; $[\theta -\Delta \theta/2, \theta +\Delta \theta/2)$. 
%DK: Do we need to include in the equation a Delta Omega for the size of the viewing angle bin?
%Gesa: Indeed, I have corrected the equation. 

In the DCT method, the signal-to-noise ratio (S/N) values of the synthetic spectra depend on the number of OPs simulated and the choice of the bins of arrival time, frequency, and viewing direction. 
Therefore, a higher resolution in viewing angle results in fewer OPs per bin and a lower S/N.  The DCT has been the default implementation in the SEDONA code. 

\subsection{Virtual Packets}

%\cite{Lucy1999FormalIntegral} discussed a formal integral method for improving the  S/N  in a MCRT simulation. 
%Extensions of this method have been introduced in the multi-dimensional simulations of active galactic nuclei (e.g. \cite{Sim2010AGN}) and stellar accretion disk winds (e.g. \cite{Long2002AccrDisk}). 
%DK: I commented out the above because it wasn't clear to me that this was relevant to the VP method that you are interested in. 
%Gesa: According to Bulla et al 2015, Prof. Mattia Bulla is inspired by these old papers to have the VP idea, and these papers indeed have used the algorithm of virtual packets but did not make specific definition of 'virtual packets'. I am not sure if I will also make mention of these papers. 
Several techniques have been suggested to improve the S/N in MCRT simulations. 
The idea of a ``virtual packet" (VP) in the SN simulation context is discussed by \cite{Kerzendorf2014Tardis} and implemented in the 3-D simulation by \cite{Bulla2015VPack}. 
Generally, VPs are used as follows
\begin{itemize}
    \item An observer viewing direction is specified at the start of the MCRT simulation. 
    \item VPs are created in the SN plasma to represent the emitted and the scattered photons at the corresponding volume-time-frequency bin pointing to the viewing direction. 
    %DK: Would it be correct to say instead that "VPs are created at each emission and scattering event to represent photons that propagate towards the observer. The energy of each VP is weighted by the probabilty that the emission/scattering event will lead to photon emission along that direction"
    %Gesa: It may not be appropriate. This statement is only true for EBT, while there are a few more algorithms also uses the concept of VPs, like the TARDIS paper, and trajectory-based technique (TBT, the other algorithm in Bulla et al. 2015). 
    \item The energy of each VP is reduced by the optical depth along the path to the observer.  
    \item Finally, the escaped VPs are used to synthesize the spectrum for the specified observer viewing angle. 
\end{itemize}
%By forcing VPs to be produced along the specified observer viewing direction, EBT reduces the S/N relative to DCT (for which only those packets that escape along the observer direction contribute to the synthetic spectrum).  
%Among the two VP methods proposed in \cite{Bulla2015VPack}, the event-based technique (EBT) provides a good balance between the computation time and the result S/N. 
%DK: I commented out the above sentence since you haven't specified what the two different methods used in Bulla are. 
%Gesa: Okay. 
Event-based techinque (EBT), proposed by \citet{Bulla2015VPack}, utilizes VPs to increase the S/N relative to DCT. 
In the EBT calculation, when an OP has undergone a physical event, a VP is created at the same coordinate and time with the same co-moving frame frequency. 
If the interaction event is Thomson scattering, then the co-moving frame Stokes vector of the VP is calculated from the Rayleigh scattering rule \citep{Chandra1960RT}: 

\begin{equation}\label{eq:EBTThomScat}
    \Bar{\boldsymbol{S}}_{VP}=\Bar{\boldsymbol{S}}_{OP} \boldsymbol{P}(\Bar{\theta}_{OP},\Bar{\varphi}_{OP},\Bar{\theta},\Bar{\varphi})\ , 
\end{equation}
where $\Bar{\boldsymbol{S}}_{OP}$ is the Stokes vector of the OP in the co-moving frame, $(\Bar{\theta}_{OP},\Bar{\varphi}_{OP})$ is the OP propagating direction in the co-moving frame, $(\Bar{\theta},\Bar{\varphi})$ is the viewing direction in the co-moving frame, $\boldsymbol{P}$ is the combination of the rotation matrix and the scattering matrix (See Equation 10 in \cite{Bulla2015VPack}). 
In the following, co-moving frame quantities are denoted with a bar, whereas non-bar quantities refer to the lab frame. 
A detailed expression of the $\boldsymbol{P}$ matrix is given in Appendix \ref{app:scatmat}. 

If the interaction event is a bound-bound transition, bound-free transition, or free-free transition, then an depolarized VP propagating towards the viewing direction is created. 
The energy of the VP is
%DK: Does the frequency of the VP change if the event is bound-free or free-free, since the photon is absorbed and re-emitted at a different frequency?
%Gesa: If a VP is created due to bound-free or free-free transition, then when the transition is done, the co-moving frame frequency of the corresponding OP is also the co-moving frame frequency of this created VP. After the creation, the lab frame frequency of a VP will never change. 
\begin{equation}
    \Bar{E}_{VP}=\frac{1}{4\pi}\Bar{E}_{OP}\ . 
\end{equation}
%DK: It seems you are using a bar to indicate comoving frame quantities, but I don't recall you stating that this is the case. 
%Gesa: It was mentioned in the footnote, just before equation 4. I have moved this to the main paragraph. 
After creation, the VPs propagate along the viewing direction with the speed of light, and are not considered in the subsequent computation of the plasma excitation and ionization states. 
The total optical depth along the VP trajectory is integrated as

\begin{equation}\label{eq:vpTau}
    \tau_{tot}=\sum_{j}d_j \alpha_{tot,j}\ , 
\end{equation}
where $d_j$ is the length of the $j$-th line segment of the VP trajectory, $\alpha_{tot,j}$ is the total extinction coefficient at the $j$-th line segment. 
The lab frame extinction coefficient is calculated from the co-moving frame value 

\begin{equation}\label{eq:opacLTrans}
    \alpha_{tot,j}(\nu)=(\Bar{\nu}/\nu)\Bar{\alpha}_{tot,j}(\Bar{\nu})\ , 
\end{equation}
where $\Bar{\nu}$ and $\nu$ are the co-moving frame frequency and the rest-frame frequency respectively. 
%DK: I'm not sure what you mean by "the line segment is split". Do you mean that the length of any individual line segment is determined by taking the smallest of these lengths? 
%Gesa: Yes, and using the following criterion, we can numerically assume the extinction coefficient inside an individual line segment is not changing, and therefore the integral of the optical depth can be calculated by the summation over the line segments. 
%DK2: I'm not sure that you defined \bar{\nu}
%Gesa2: I have added the definition. 

The VP is moved in small line segments, the length of which are determined by finding the minimum distance among the following
\begin{itemize}
    \item The VP reaches the boundary of a volume cell. 
    \item The VP reaches the end of a time step. 
    \item The co-moving frame frequency of the VP reaches the boundary of the frequency grid of $\alpha_{tot}$. 
\end{itemize}

The final spectrum is constructed by summing the VPs 
%DK: Don't all VPs escape, since you haven't specified a mechanism by which they destroyed? So should you say that the final spectrum is constructed by summing  all VPs with their energy attenuated due to the optical depth to the observer
%Gesa: Yes and no. There is no mechanism in this paper to destroy a VP, and all the VPs will escape eventually. In Bulla et al 2015, there is a mechanism to destroy the VPs. When the accumulated optical depth of a VP reaches 10, then the VP is destroyed. For the consistency of the paper, I have removed the description of 'escape'. 

\begin{equation}\label{eq:sumVP}
    \begin{pmatrix} \mathcal{I} \\ \mathcal{Q} \\ \mathcal{U} \\ \mathcal{V} \end{pmatrix} = \frac{1}{4\pi r^2 \Delta t \Delta \nu} \sum_i E_{VP,i} \boldsymbol{s}_{VP,i} e^{-\tau_{tot,i}}\ , 
\end{equation}
the summation is over all the VPs in the frequency bin $[\nu-\Delta \nu/2,\nu+\Delta \nu/2)$, and arrival time interval $[t-\Delta t/2, t+\Delta t/2)$. 

%DK: Perhaps the paragraph below could be rewritten as:
% "The EBT can be accelerated by only emitting VPs with frequencies within the desired spectral range of the synthetic observables, and  by removing VPs once the integrated optical depth becomes large, $\tau_{\tot} \ge 10$. We don't apply these optimizations in the present calculations, but Bulla find that they can decrease the computational time by a factor of $\sim 4$ while not affecting accuracy in most case . 
%Gesa: I have changed the text. 

The EBT can be accelerated by only emitting VPs with frequencies within a desired  spectral range (e.g., $3500-10000\ \AA$) for the synthetic observables, and by removing the VP once the integrated optical depth reaches a critical value $\tau_{tot} = 10$. 
We do not apply these optimizations in the present calculations, but \cite{Bulla2015VPack} find that they can decrease the computational time by a factor of $\sim 4$ while not affecting accuracy. 

%We implement the EBT into SEDONA. 
%Comparing to the EBT recipe used in ARTIS \citep{Bulla2015VPack}, we did not include two features: 

%\begin{itemize}
%    \item The VP is deleted when its $\tau_{tot}$ reaches 10. 
%    \item The VP is not created if its rest-frame wavelength will not be in the $3500-10000\ \rm\AA$ range. 
%\end{itemize}

%According to \cite{Bulla2015VPack}, these two features could increase the computation speed by a factor of $\sim 4$, and the $\tau_{tot}<10$ criterion does not affect the accuracy in most of the cases. 

\section{The Integral-Based Technique}\label{sec:ibt}

Our proposed IBT uses integrated packets (IPs), which are an improved version of VPs that can be used to more efficiently construct spectra. 
Compared to VP, the major changes of IP are: 
(1) A VP stores the energy and polarization at a single frequency with a four-vector $\boldsymbol{S}_{VP}$, while an IP stores the energy and polarization at multiple frequency bins with a $4\times N$ tensor of cell radiance $\boldsymbol{R}([I,Q,U,V];[\nu_1,...,\nu_N])$; 
(2) The creation of a VP is based on the physical interactions of OPs, whereas an IP is created in each volume cell for each time step. 

%DK: You may want to explain briefly here how IPs differ from VPs. 
%Details are illustrated in the following subsections. 
%Gesa: I have listed the major differences of IP from VP. 
\subsection{The Integrated Packets}

%In a 3-D time-dependent MCRT calculation, the SN structure is divided into several volume cells with volume $\Delta V$, the time step is $\Delta t$, the polar coordinate of the viewing direction is at ($\varphi$,$\theta$). 
The IP computation is based on a universal logarithmically spaced frequency grid $[\nu_0,\nu_1...\nu_N]$. 
The frequency ratio between the adjacent pixels is a constant $C$, satisfying 
\begin{equation}\label{eq:freqGrid}
    \nu_{n+1}/\nu_n=C. 
\end{equation}
%DK: This leads to a logarthimically spaced grid, correct? It might be nice to say that explicitly.  Does the method require a logarthimic spaced grid or could it be applied to other grids? 
%Gesa: Yes, it is a logarithmically spaced grid. It may be possible to use other frequency grids or even heterogeneous frequency grids when doppler effect is not strong. However, in the specific scenario of supernova, it is better to use logarithmically spaced grid because this grid structure simplifies the computation of doppler effect to the shift of grids. 
Using this frequency grid, the Doppler shift of the variables could be simplified to the shifting of the index of the frequency bin, which saves computation time. 
%DK: What do you mean by "pixel"? I think it means shifting the index of the frequency bin.
%Gesa: Yes, that is what I mean. 

%DK: What do you mean by "radiation energy" in the paragraph below? What are the units of R?  And why are you introducing it here? 
%Gesa: I have added the unit of radiation energy (cell radiance) here, and added more discussion at the end of this subsection to clarify the usage of radiation energy (cell radiance) in the program. 
% Lifan: \textcolor{red}{The radiation energy is a misleading name as the quantity does not have the units of energy. }
We define the cell radiance $\boldsymbol{R}(\Vec{x},t,\nu)$ in the lab frame for a cell located at coordinate $\Vec{x}$ and time $t$ as the energy radiated per frequency bin at frequency $\nu$ toward the viewing direction, the units of $\boldsymbol{R}(\Vec{x},t,\nu)$ is $\rm erg\ Hz^{-1}\ sr^{-1} $. 
%, and frequency $\nu$ toward the viewing direction. 
The formal solution of the cell radiance is: 
%The integrated Stokes vector of the volume cell ($\Vec{x}$) at the time step ($t$) at the wavelength $\nu_n$ is calculated as. 
%\begin{widetext}
%    \begin{equation}\label{eq:RadEng}
%        \boldsymbol{R}(\Vec{x},t,\nu) =\Delta v \Delta t \left(\frac{\Bar{\nu}}{\nu}\right)^{-2} \left( \Bar{j}_{emi}(\Vec{x},t,\Bar{\nu}) \begin{pmatrix}1\\0\\0\\0\end{pmatrix} + \alpha_T \int_{\Omega} \Bar{\boldsymbol{I}}(\Vec{x},t,\Bar{\theta}_{in},\Bar{\varphi}_{in},\Bar{\nu}) \boldsymbol{P}(\Bar{\theta}_{in},\Bar{\varphi}_{in},\Bar{\theta},\Bar{\varphi}) d\Omega \right)\ , 
%    \end{equation}
%\end{widetext}

\begin{eqnarray}\label{eq:RadEng}
    \boldsymbol{R}(\Vec{x},t,\nu) =\Delta v \Delta t_s \left(\frac{\Bar{\nu}}{\nu}\right)^{-2} \left(\Bar{\boldsymbol{j}}_{emi}+\Bar{\boldsymbol{j}}_{sc}\right) \\ 
    \Bar{\boldsymbol{j}}_{emi} =\Bar{j}_{emi}(\Vec{x},t,\Bar{\nu}) \begin{pmatrix}1\\0\\0\\0\end{pmatrix} \\ 
    \Bar{\boldsymbol{j}}_{sc} =\alpha_T\int_{\Omega} \Bar{\boldsymbol{I}}(\Vec{x},t,\Bar{\theta}_{in},\Bar{\varphi}_{in},\Bar{\nu}) \boldsymbol{P}(\Bar{\theta}_{in},\Bar{\varphi}_{in},\Bar{\theta},\Bar{\varphi}) d\Omega, 
\end{eqnarray}
where $\Bar{\boldsymbol{I}}$ is the specific intensity as a four-vector of the Stokes parameters; 
$\Delta t_s$ is the size of the simulation time step; 
$\Bar{\boldsymbol{j}}_{emi}$ and $\Bar{\boldsymbol{j}}_{sc}$ are the four-vectors of the emission terms due to depolarized isotropic emission and scattered light from Thomson scattering, respectively; 
the $\Bar{j}_{emi}$ function is the depolarized isotropic emission term, which includes all the emission from bound-bound, bound-free, and free-free transitions; 
the $\boldsymbol{P}(\Bar{\theta}_{in},\Bar{\varphi}_{in},\Bar{\theta},\Bar{\varphi})$ function is a combination of the Rayleigh scattering phase matrix and rotation matrix discussed in Appendix \ref{app:scatmat}; 
$\alpha_T$ is the extinction coefficient for Thomson scattering; 
%($\Bar{\theta},\Bar{\varphi}$) is the viewing direction converted to the co-moving frame; 
the integral is over all the incident directions ($\Bar{\theta}_{in},\Bar{\varphi}_{in}$); the $(\Bar{\nu}/\nu)^{-2}$ term is the correction of the Doppler effect (\citealt[e.g.,][eq.~(1--3)]{Castor1972RTE}; see also \citealt[][p.~31,33,495--496]{Mihalas1978Book}). 

For each IP, the cell radiance $\boldsymbol{R}$ is calculated on the universal frequency grid and stored as a $4\times N$ vector. 
During the propagation of an IP, $\boldsymbol{R}$ is reduced by the extinction coefficient on the trajectory. 
The spectral synthesis is the summation of $\boldsymbol{R}$ over the IPs at a specific arrival time. 
Section \ref{subsec:isoemission} and Section \ref{subsec:pathintegral} illustrate the computation of $\boldsymbol{R}$ in a MCRT simulation during the creation of an IP. 
Section \ref{subsec:propagation} illustrates the update of $\boldsymbol{R}$ during the propagation of IP and spectral synthesis. 

\subsection{Isotropic Emission}\label{subsec:isoemission}

The isotropic emission term, including bound-bound, bound-free, and free-free interactions, can be calculated by the summation of the OPs from the same physical events 

\begin{equation}
    \Delta v \Delta t_s \left(\frac{\Bar{\nu}}{\nu}\right)^{-2} \Bar{j}_{emi}(\Vec{x},t,\Bar{\nu})=\left(\frac{\Bar{\nu}}{\nu}\right)^{-2}\sum_i \frac{\Bar{E}_{OP,i}}{4\pi \Delta \Bar{\nu}}\ , 
\end{equation}
where the summation is over all the OPs within the frequency bin $[\Bar{\nu}-\Delta \Bar{\nu}/2, \Bar{\nu}+\Delta \Bar{\nu}/2)$, the volume cell $\Delta v$, and the time step bin $[t_s,t_s+\Delta t_s)$. 
Only OPs that are newly created or undergone the same physical events are taken into account. 
%DK: Above you said that the IBT method usings IPs instead of VPs. So should these be called IPs? You haven't specified yet what is different about IPs and VPs. 
%Note that the VPs created in this step are not propagated in the following calculations, and the computation resources are recycled. 
%This recipe is suitable to complicated physical events with partial frequency redistribution (i.e. macroatom \cite{Lucy2002macroatom}). 
%DK: I don't know what you mean by "computation resources are recycled", or what the "recipe" is you are talking about.
%Gesa: There is no need to make mention of VPs here, therefore I have changed E_VP to E_OP/4/pi so that the equation is still the same. 
%Gesa: When the emission term is explicitly available, just like the line expansion opacity calculation in SEDONA, then there is no need to use the above paragraph. However, if the emission terms is not explicitly available, just like the "macroatom" calculation in Lucy 2002, the above paragraph can be used. 

In the present SEDONA calculation, we include bound-free and free-free transitions and adopt the line expansion opacity approximation (\citealt[][eq.~(8)]{Kasen2006Sedona}) for bound-bound transitions which are treated as purely absorptive. 
Therefore, the isotropic emission term is calculated from the current plasma state as $\Bar{j}_{emi}(\Vec{x},t,\Bar{\nu})=B(\Bar{\nu})\Bar{\alpha}_{abs}(\Vec{x},t,\Bar{\nu})$, where $B(\Bar{\nu})$ is the blackbody spectrum and $\Bar{\alpha}_{abs}$ is the total extinction coefficient including bound-bound, bound-free, and free-free transitions (Thomson scattering is not included in this term). 
%This recipe generates an emission term without Monte-Carlo noise, which accelerates the code by avoiding the time consuming summation computation. 
%DK: What is the time-consuming summation computation you refer to? 
%Gesa: One can use the summation equation (Eq. 11) to calculate the isotropic emission term of the radiation energy (cell radiance). However, Eq 11 is subject to the number of optical packets (which could introduce extra computation time due to a large number of packets, or introduce extra MC noice due to a small number of packets). While In SEDONA with line expansion opacity option, the $\Bar{j}_{emi}$ in the left hand side of Eq. 11 is already available in SEDONA under the sobolev approximation, so there is no need to use the right hand side of Eq. 11. 

\subsection{The Path Integral}\label{subsec:pathintegral}
%DK: This disussion of the IPs in this and the next section are crucial for explaining the new idea, but I couldn't quite follow what was happening. How exactly are IPs created (what is their energy and frequency?)   
%Gesa: As the updated description of IP above, the radiation energy (cell radiance) is a 4*N tensor to store the energy and polarization at different frequency bins. This subsection and the previous subsection illustrates how the radiation energy (cell radiance) is calculated during the creation of an IP. 
The cell radiance from Thomson scattering is calculated from the path integral of the OPs 

%\begin{widetext}
%    \begin{equation}\label{eq:IBTThomScat}
%        \Delta v \Delta t_s \left(\frac{\Bar{\nu}}{\nu}\right)^{-2} \sigma_T \int_{\Omega} \Bar{\boldsymbol{I}}(\Vec{x},t,\Bar{\theta}_{in},\Bar{\varphi}_{in},\Bar{\nu}) \boldsymbol{P}(\Bar{\theta}_{in},\Bar{\varphi}_{in},\Bar{\theta},\Bar{\varphi}) d\Omega =  \frac{1}{4\pi\Delta \Bar{\nu}}\sum_{i,j} \bar{E}_i \bar{d}_{i,j} \Bar{\alpha}_{T}\Bar{\boldsymbol{s}}_{i,j} \boldsymbol{P}(\Bar{\theta}_{i,j},\Bar{\varphi}_{i,j},\Bar{\theta},\Bar{\varphi})
%    \end{equation}
%\end{widetext}

\begin{equation}\label{eq:IBTThomScat}
    \Delta v \Delta t_s \left(\frac{\Bar{\nu}}{\nu}\right)^{-2} \Bar{\boldsymbol{j}}_{sc} = \frac{\sum_{i,j} \bar{E}_i \bar{d}_{i,j} \Bar{\alpha}_{T}\Bar{\boldsymbol{s}}_{i,j} \boldsymbol{P}(\Bar{\theta}_{i,j},\Bar{\varphi}_{i,j},\Bar{\theta},\Bar{\varphi})}{4\pi\Delta \Bar{\nu}}
\end{equation}
where $\bar{d}_{i,j}$ is the co-moving frame length of the $j$-th line segment of the $i$-th OP trajectory in the volume cell and in the time step, the summation is over all the trajectories within the frequency bin $[\Bar{\nu}-\Delta \Bar{\nu}/2, \Bar{\nu}+\Delta \Bar{\nu}/2)$, the volume cell $\Delta v$, the time step bin $[t_s,t_s+\Delta t_s)$. 
%DK2: It is hard to follow the text with this long equation breaking up the top and bottom. I'd recommend defining a new quantity (e.g., j_scat) that is defined as this integral over I and P. Then this equation and Eq 10 may both fit into one column, which would be much more readable.
%Gesa2: I have modified the two equations, they are fit into one column. 

For each volume cell and at each time step, an IP is created. 
The initial coordinate of the IP is randomly chosen in the volume cell, and the initial time of the IP is randomly chosen in the time step. 
The cell radiance $\boldsymbol{R}$ is calculated on the universal frequency grid and stored in IP as a $4\times N$ tensor. 
%DK: What do you mean the frequency grid is stored in each IP.  Does this mean each IP is actually a whole spectrum ? 
%DK: DO you really need to emit an IP from each cell? Cells near the outer edge of the grid will have low density and emission and may not contribute much at all. While cells deep inside the ejecta may be very optically thick and radiation may not escape from them.
%Gesa: Yes, the radiation energy (cell radiance) of an IP is effectively the spectrum of the corresponding plasma cell, the supernova spectrum is the summation of the radiation energy (cell radiance). 
%Gesa: Actually in my SEDONA calculation, I have ignored the cells with a temperature equal to or smaller than 1000 K, therefore the unimportant cells at outer layers are ignored to increase the computation speed. 

\subsection{The Propagation of Integral Packets}\label{subsec:propagation}

After creation, the IPs propagate toward the viewer with the speed of light, and the cell radiance $\boldsymbol{R}$ is reduced during propagation. 
Similar to the propagation of the VPs, the trajectory of an IP is split into line segments by the edges of time steps and volume cells. 
Moreover, the trajectory is further split to satisfy the criterion: 

\begin{equation}\label{eq:criterion}
    \left|\frac{\Bar{\nu}_{begin}-\Bar{\nu}_{end}}{\Bar{\nu}_{end}} \right| \leq C-1  , \ 
\end{equation}
where $\Bar{\nu}_{begin}$ is the co-moving frame frequency at the beginning of the line segment, $\Bar{\nu}_{end}$ is the co-moving frame frequency of the same lab frame frequency $\nu$ at the ending of the line segment, the constant $C$ is defined in Equation \ref{eq:freqGrid}. 

In the calculation of VP in Equation \ref{eq:vpTau} and \ref{eq:sumVP}, the total opacity is accumulated over the trajectory, which computationally requires one more double precision floating point number per VP to store the opacity. 
This process reduces the computation time on exponentials at the cost of memory space. 
While in the propagation of IPs, the cell radiance is updated at every line segment on the trajectory without accumulating the total opacity over the trajectory
%DK: I'm not sure what is meant by R_new and R_old in the equation below
\begin{equation}\label{eq:updateRad}
    \boldsymbol{R}_{i,j+1}(\nu)=\boldsymbol{R}_{i,j}(\nu) e^{-d_{i,j} \alpha_{tot,i,j}(\nu)}\ , 
\end{equation}
where $d_{i,j}$ is the length of the $i$-th IP at the $j$-th line segment over the trajectory, $\alpha_{tot,i,j}(\nu)$ is the total extinction coefficient of the $i$-th IP at the $j$-th line segment and at lab frame frequency $\nu$. 
%Note that in SEDONA, $\Bar{\alpha}_{tot,j}$ being calculated over the frequency grid in the co-moving frame as an array of $N$ elements allows the . 
The final spectrum is the sum of the IPs: 

\begin{equation}\label{eq:sumIP}
    \begin{pmatrix} \mathcal{I} \\ \mathcal{Q} \\ \mathcal{U} \\ \mathcal{V} \end{pmatrix} = \frac{1}{4\pi r^2 \Delta t } \sum_i \boldsymbol{R}_{i,final}\ , 
\end{equation}
where $\boldsymbol{R}_{i,final}$ is the final cell radiance when the $i$-th IP escapes the simulation domain. 
The summation is over all IPs in the arrival time interval $[t-\Delta t/2, t+\Delta t/2)$. 
Note that the grid size of arrival time $\Delta t$ should be much larger than the time step size $\Delta t_s$ to avoid grid mismatch issue, while this issue is not significant in DCT or EBT calculations. 

\section{Model Validation}\label{sec:validation}

\begin{figure*}[htb!]
    \centering
    \includegraphics[width=\textwidth]{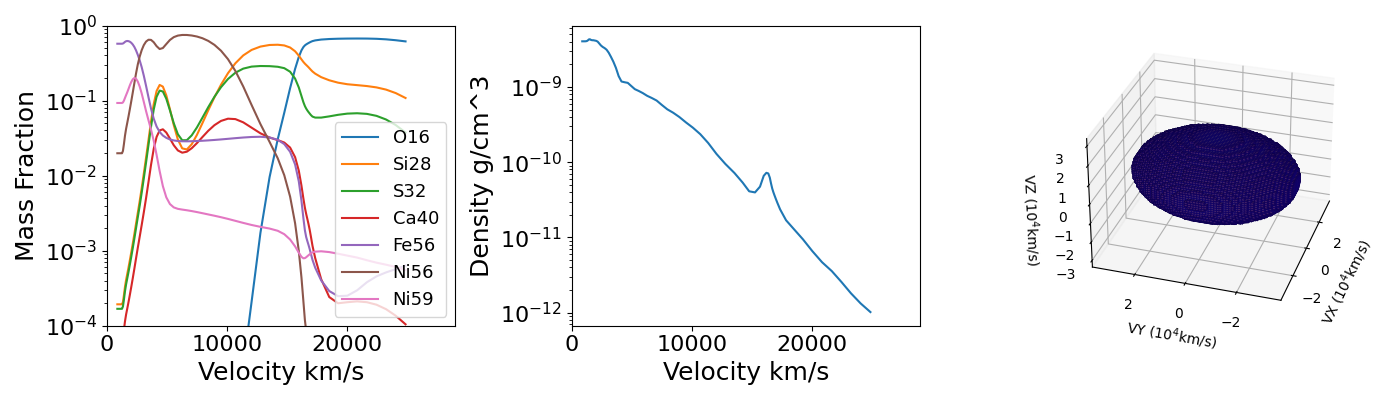}
    \caption{Left panel: the mass fractions of several isotopes in the DDC10 model. 
    Middle panel: the density structure of the DDC10 model. 
    Homologous expansion is assumed, so the radius is equal to velocity times time. 
    Right panel: the geometry of the modified DDC10 model with 3-D structures (D3D model). 
    %, the viewing direction is $cos(\theta) = 0.5$, $\varphi = 3.4455$. 
    }
    \label{fig:ddc10}
\end{figure*}

In this section, we prepare two toy models to validate the IBT computation results. 
The first model is the SN~Ia delayed-detonation model DDC10. 
DDC10 is calculated from hydrodynamic simulations in \cite{Blondin2013DDC10} and adopted as a benchmark in \cite{Blondin2022Standart} to evaluate the differences between radiative transfer codes. 
We remap the 1-D model into a $60\times 60\times 60$ 3-D Cartesian grid. 
retaining the spherical symmetry and with the outer boundary velocity limited to 25,000 km/s. 
Figure \ref{fig:ddc10} shows the abundance of several elements and the density profile of the DDC10 model. 
The second model is an ellipsoidal model based on the DDC10 model; the y and z coordinates are reprojected with $y_{new}=1.3 y$ and $z_{new}=z/1.3$, in order to introduce asphericity. 
%DK: I believe this is an ellipsoidal model?  Is it correct to say that the density contours are made ellipsoidal with axis ratio of 1/3 
%Gesa: yes, it is an ellipsoiday model. 
This model is also realized on a $60\times 60\times 60$ Cartesian grid, with a boundary velocity of 32,500 km/s. 
The geometry of the model is shown in the right panel of Figure \ref{fig:ddc10}. 

In the following text, the spherical symmetric DDC10 model is denoted as D1D, and the modified ellipsoidal 3-D model is denoted as D3D. 
Both the models are treated with 3-D time-dependent radiative transfer calculations in SEDONA starting from 0.976 days after the explosion. 
Before 6.6 days, the simulation time step is a logarithmic with $\Delta t_s/t_s=0.03$. 
After 6.6 days, the simulation time step is $\Delta t_s = 0.2 $ days. 
The arrival time bin size is $\Delta t = 1$ day, which satisfies $\Delta t \geq 5 \Delta t_s$. 
The frequency grid ranges from $1\times 10^{14}$ Hz to $5\times 10^{15}$ Hz (30,000 $\AA$ to 600 $\AA$) with $\Delta \nu / \nu=0.002$. 

\subsection{Spherical Symmetry}

\begin{figure*}
    \centering
    \includegraphics[width=0.49\textwidth]{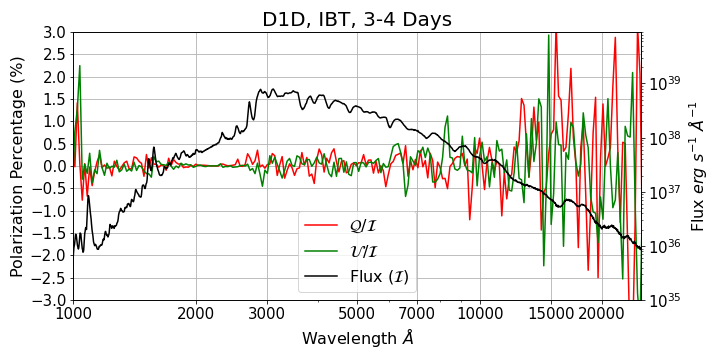}
    \includegraphics[width=0.49\textwidth]{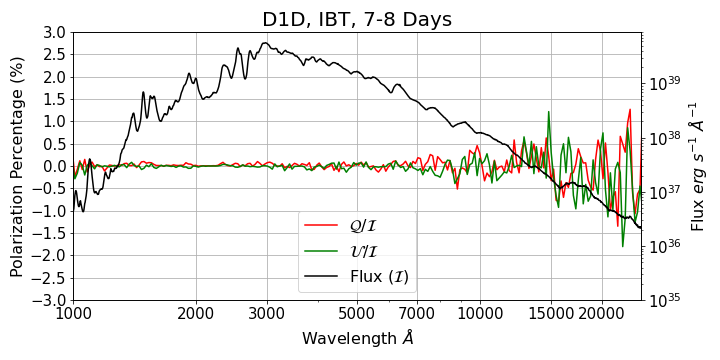}
    \includegraphics[width=0.49\textwidth]{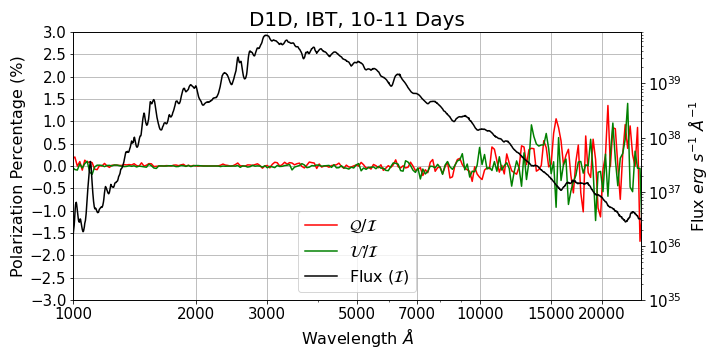}
    \includegraphics[width=0.49\textwidth]{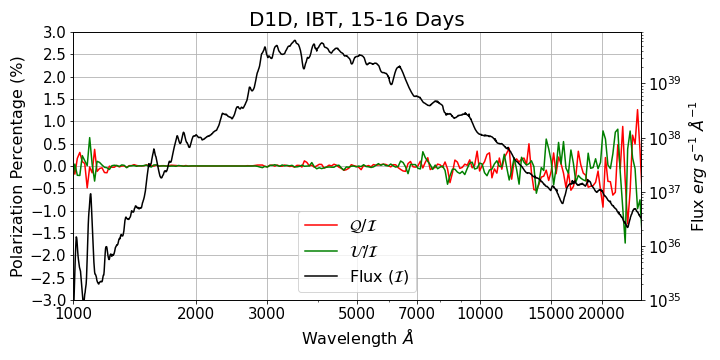}
    \includegraphics[width=0.49\textwidth]{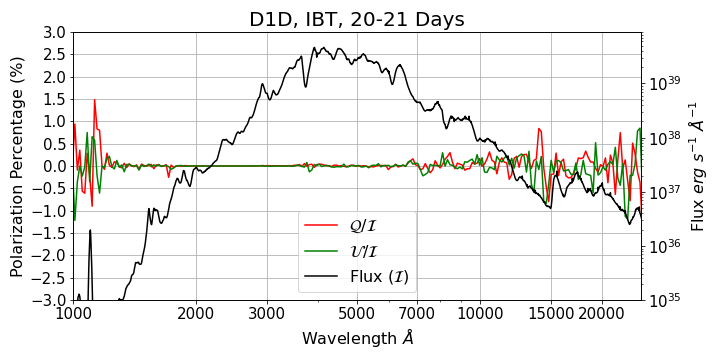}
    \includegraphics[width=0.49\textwidth]{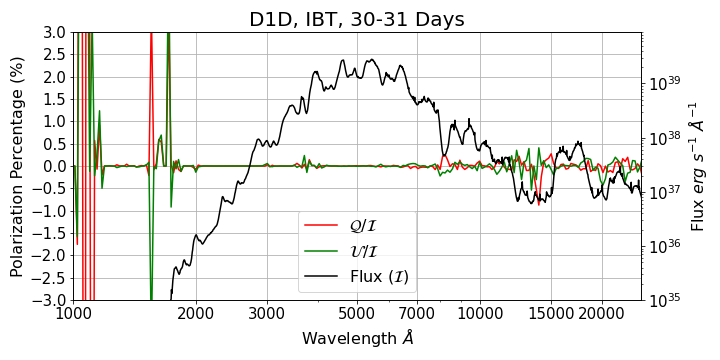}
    \caption{The spectral flux ($\mathcal{I}$) and the polarization percentage ($\mathcal{Q}/\mathcal{I}$, $\mathcal{U}/\mathcal{I}$) of the spherical symmetric D1D model at different times. 
    The spectra are calculated by IBT with $6\times 10^{6}$ OPs per simulation time step. 
    The arrival time bin relative to the explosion time is labeled in the title of each subplots. 
    The polarization percentages of $\mathcal{Q}/\mathcal{I}$ and $\mathcal{U}/\mathcal{I}$ are averaged using a 7-pixel bin. }
    \label{fig:sphere}
\end{figure*}

Because the linear polarization components cancel out, a spherical symmetric model should produce zero polarization. 
Using IBT, we calculate the spectropolarimetry of the D1D model to test if the algorithm recovers zero polarization. 
Figure \ref{fig:sphere} shows the spectra and the polarization percentage at different times after the explosion. 
At each time step, $6\times 10^{6}$ OPs are generated to represent the energy release from radioactive decay and gamma-ray scattering. 
We find that the polarization signal at all times and wavelengths is consistent with the expected zero polarization. 
Moreover, the simulation error between 7 days and 31 days and between 2000 and 10000 $\AA$ wavelength range (within which most SNe Ia spectropolarimetry observations are made \citep{Cikota2019PolObs}) is as low as $\sim 0.2 \%$. 
Notably, the spectra in the ultraviolet wavelength between 7 days and 21 days also show high S/N, despite the emergent flux being low in this band. 
In the early phase of SNe Ia, most of the OPs are generated below the photosphere, therefore the cell radiance $\boldsymbol{R}$ above the photosphere has low S/N. 
%DK: I am still not clear what the radiation energy R refers to. You should make it clearer when you introduce the concept above.
As a result, the spectrum shows low S/N at 3-4 days after the explosion. 
Moreover, the S/N in the infrared wavelength is lower than that in the optical wavelength, because most of the OPs are in the optical wavelength. 
Due to the limited number of OPs in the ultraviolet wavelength at late phase, the ultraviolet spectrum at 30-31 days and after is dominated by Monte-Carlo noise. 
In the IBT simulation of this paper, this phenomenon is only observed below 2000 $\rm\AA$ and can be alleviated by increasing the number of OPs. 
%We conclude this error will not affect the application of IBT on the study of optical 

\subsection{Mirror Symmetry}

\begin{figure*}
    \centering
    \includegraphics[width=0.49\textwidth]{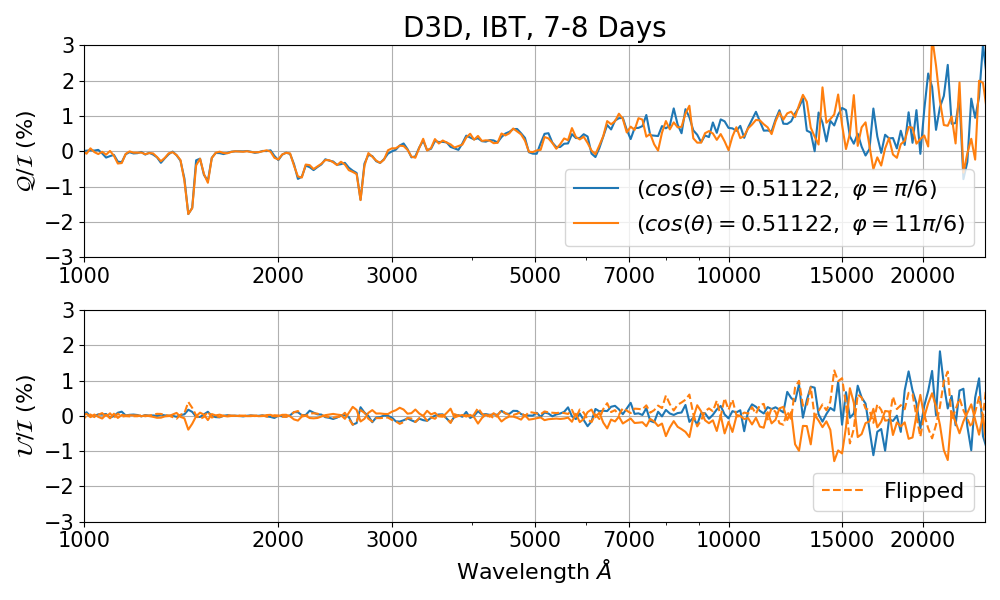}
    \includegraphics[width=0.49\textwidth]{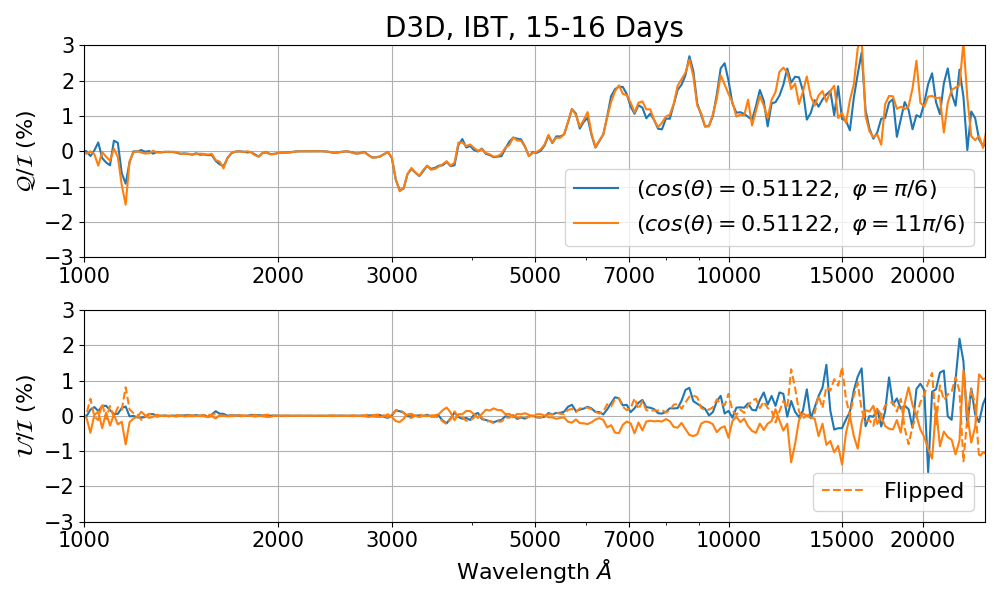}
    \includegraphics[width=0.49\textwidth]{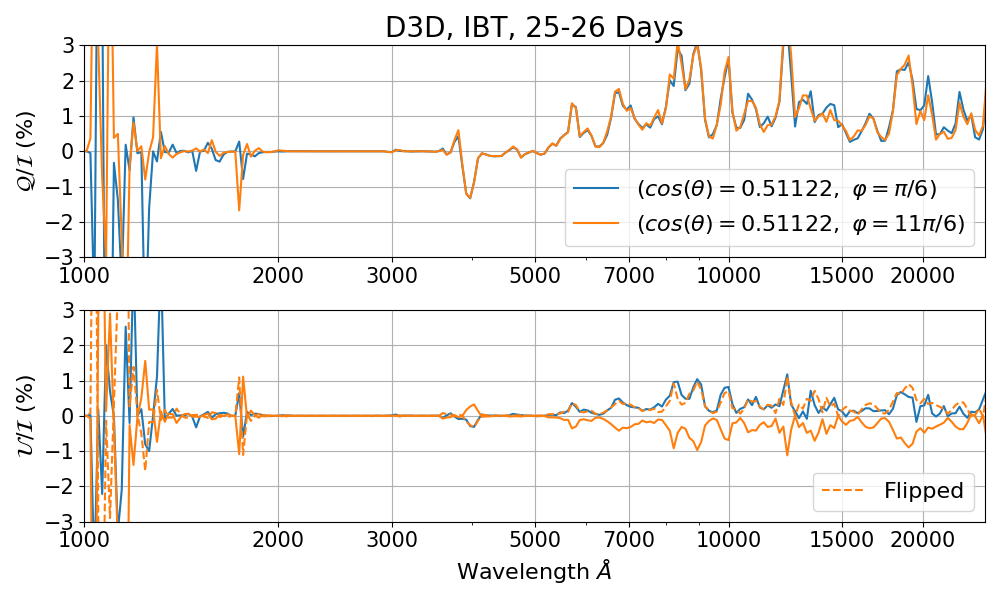}
    \includegraphics[width=0.49\textwidth]{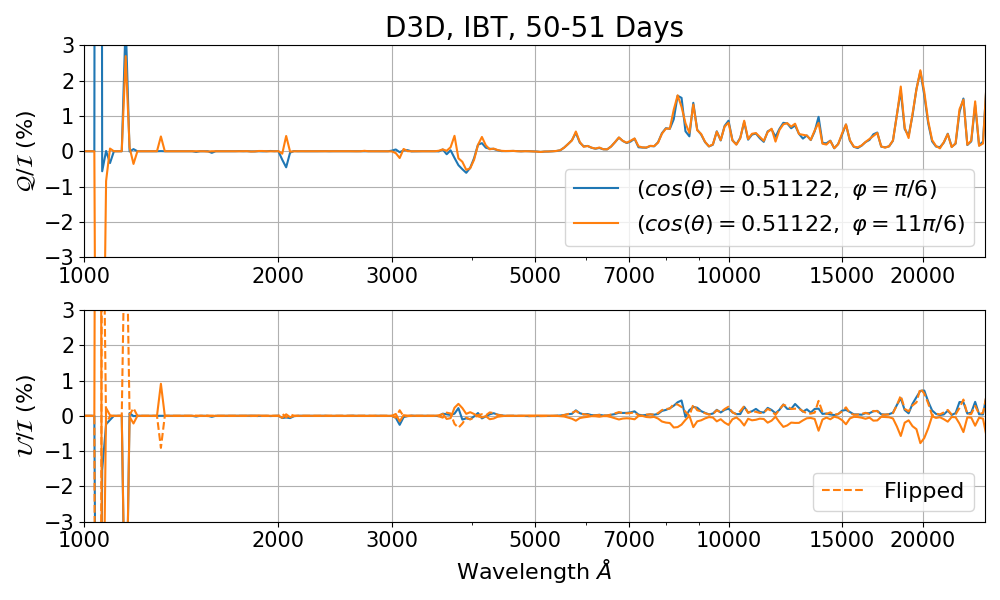}
    \caption{The spectropolarimetry of the D3D model at the two symmetric viewing directions $(cos(\theta)=0.51122,\ \varphi=\pi/6)$ and $(cos(\theta)=0.51122,\ \varphi=11\pi/6)$. 
    The arrival time bin at each panel is labeled in the title of figures. 
    In the sub-figures of Stokes $\mathcal{U}/\mathcal{I}$ term, the flipped values at one of the viewer points is shown as dashed line, in order to make direct comparisons. 
    The polarization percentages of $\mathcal{Q}/\mathcal{I}$ and $\mathcal{U}/\mathcal{I}$ are averaged using a 7-pixel bin. }
    \label{fig:symmetry}
\end{figure*}

In this section, we use IBT to calculate the spectropolarimetry at two mirror symmetric viewing directions of the D3D model. 
At each time step we generate $1\times 10^7$ OPs. 
Based on the mirror symmetry of the D3D model in the X-Z, Y-Z, and X-Y planes, spectropolarimetry in the viewing directions $(cos(\theta)=0.51122,\ \varphi=\pi/6)$ and $(cos(\theta)=0.51122,\ \varphi=11\pi/6)$ should have the same Stokes $\mathcal{Q}$ terms and inverted $\mathcal{U}$ terms. 
Figure \ref{fig:symmetry} shows the spectropolarimetry at the two symmetric viewing directions. 
We notice the simulation results are consistent with the theoretical expectations of the symmetry at above 2000 $\rm\AA$ wavelength from early phase (7-8 days after the explosion) to late phase (50-51 days after explosion). 
Similar to the spherical symmetric validation results in Figure \ref{fig:sphere}, the ultraviolet spectropolarimetry below 2000 $\rm\AA$ after $\sim$25 days are dominated by Monte-Carlo noise. 
Moreover, ultraviolet spectropolarimetry at 7-8 days and 15-16 days shows an exceptional high S/N. 

\subsection{Cross Comparison}\label{sec:crosscompare}

\begin{figure*}
    \centering
    \includegraphics[width=0.32\textwidth]{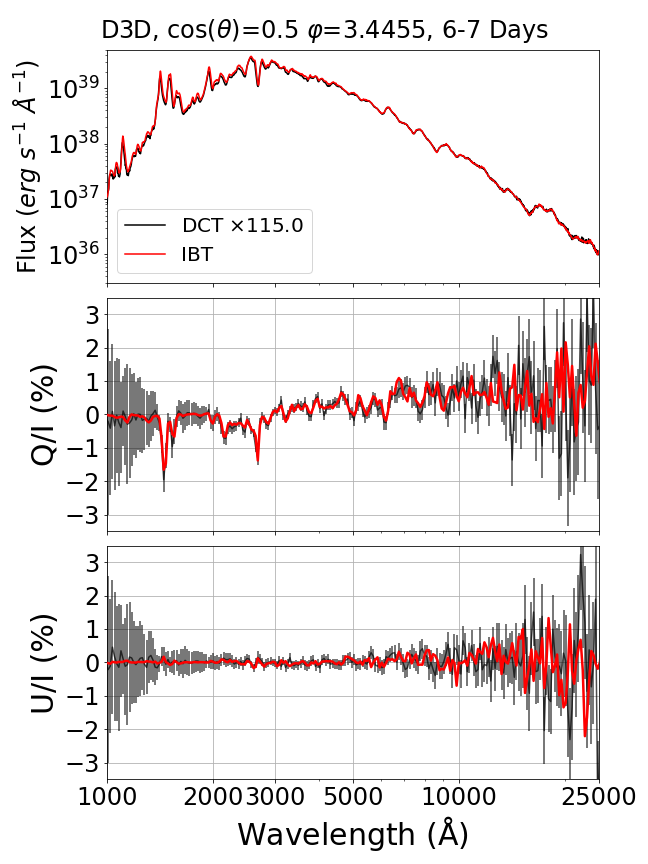}
    \includegraphics[width=0.32\textwidth]{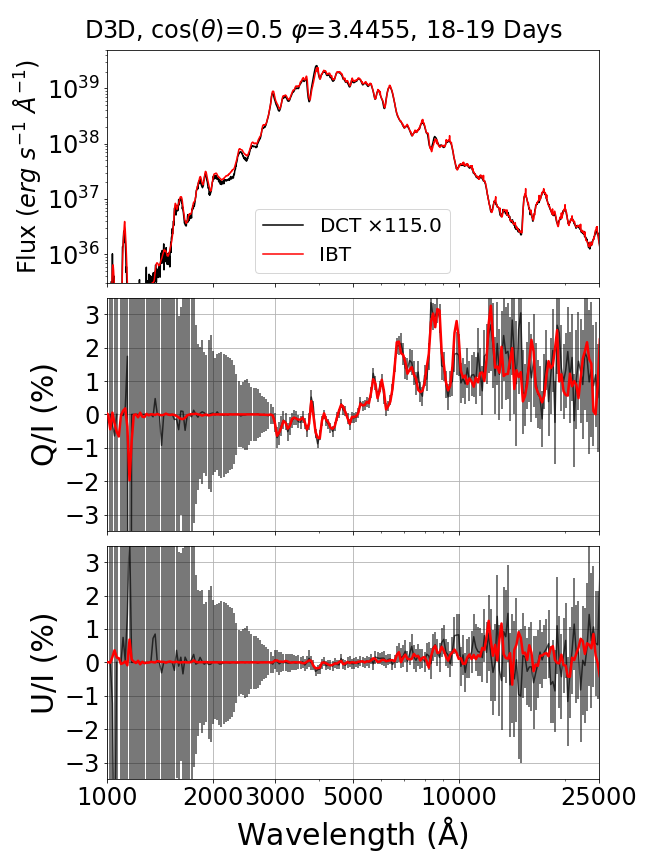}
    \includegraphics[width=0.32\textwidth]{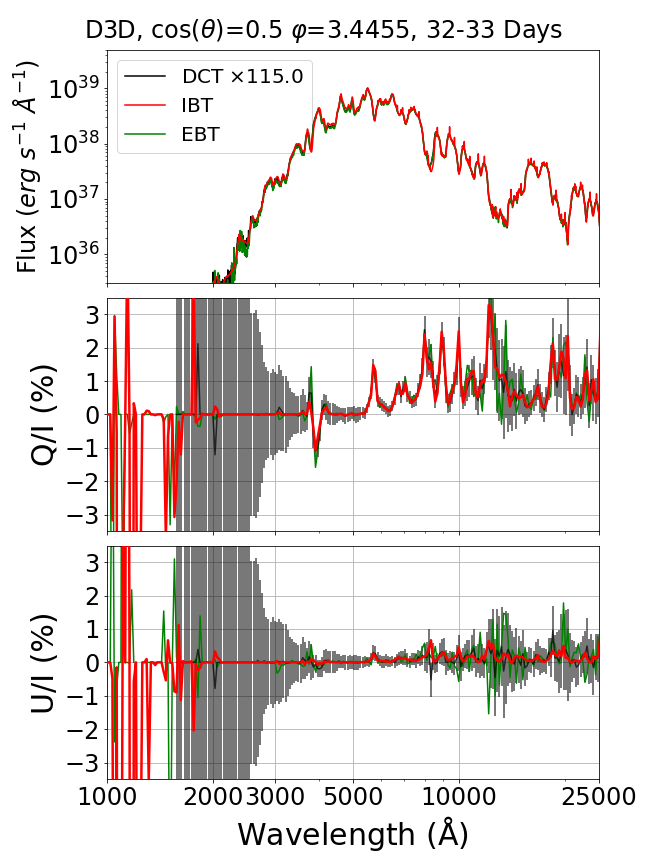}
    \caption{The spectra and linear polarization of the D3D model at the viewing direction $cos(\theta)=0.5$, $\varphi=3.4455$, using DCT (black), EBT (green), and IBT (red) methods. 
    The DCT calculation is performed with $4.6\times 10^{8}$ OPs per time step (which is $115$ times more than EBT or IBT, marked in the legend), while EBT and IBT calculation generates $4\times 10^{6}$ OPs per time step. 
    The arrival time bin is shown on top of each panel. 
    Due to the memory limit, the EBT calculation is only started at 25 days after the explosion, and the EBT results are not shown at 6-7 days and 18-19 days. 
    The Monte-Carlo error bar of the DCT result is estimated by the number of escaped OPs at each bin. 
    Due to the lack of OPs, the spectrum between 1000 \AA\ and 1500 \AA\ at 32-33 days from DCT is missing. 
    The polarization percentage is binned with 7 pixels. }\label{fig:timeseq}
\end{figure*}

In this section, we compare the simulation results from the EBT, IBT, and DCT methods using the D3D model. 
Figure \ref{fig:timeseq} shows the spectra and linear polarization time sequence calculated by the three methods. 
In the DCT calculation, the viewing direction is split into a $10\times 10$ bin and $4.6\times 10^{8}$ OPs are generated per time step to achieve a reasonable S/N. 
For the EBT and IBT calculations, we only generate $4\times 10^{6}$ OPs per time step. 
Figure \ref{fig:timeseq} shows that the IBT calculation produced spectra and polarization with a S/N comparable to a DCT calculation that used $\sim 115$ times more OPs. 
This noise reduction holds from 6 days to 33 days after the explosion and from ultraviolet to infrared wavelengths. 
%After $\sim 32$ days, the three methods are failed between 1000 and 2000 $\AA$ due to the limit of the OPs. 

Due to the high opacity at the early phase of SN, the OPs undergo multiple scattering events near the core of SN, which results in a rapidly growing number of VPs in the EBT calculation and an increasing memory usage expense. 
Therefore, in this example EBT calculation we started the calculation at 25 days after the explosion due to the limit of hardware. 

\section{Computational Performance} \label{sec:compperf}

\subsection{Computational Speed Comparison}

The computational speed of the three methods is measured on one 48-core compute node of Grace supercomputer in TAMU HPRC \footnote{https://hprc.tamu.edu/}. 
Each node has 2 sockets of Intel Xeon 6248R CPUs, in total 48 cores, and has 384 GBs of RAM memory. 
The majority of the computation time in a 3-D time-dependent radiative transfer computation consists of two components: (1) the computation of the plasma state, level population, and opacity of the grid; (2) the propagation of energy packets including OPs, VPs, or IPs. 
%DK2: I don't think you need to mention gamma-ray packets separately (since you haven't explained what they are yet) as these are basically just OPs
The choice of spectral synthesis methods only affects the propagation time of the energy packets. 

\begin{figure}[htb!]
    \centering
    \plotone{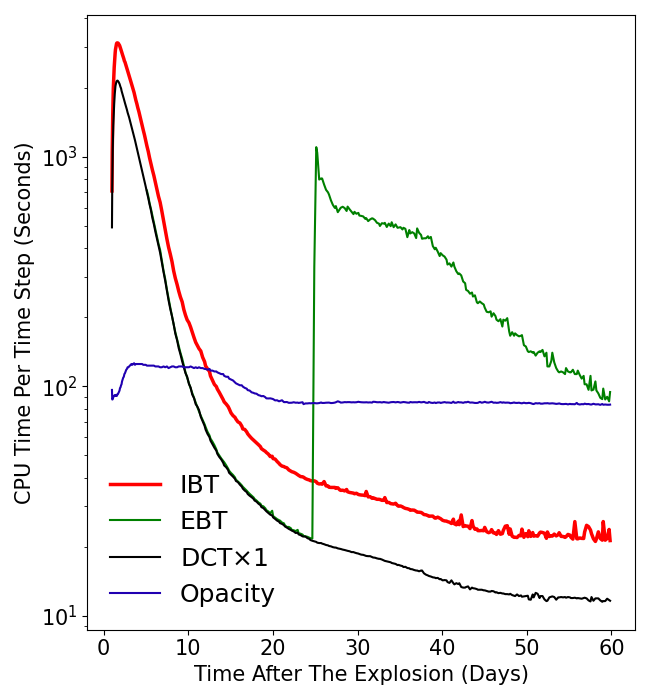}
    \caption{Comparison of particle propagation computation time at different time steps using the DCT (black), EBT (green), and IBT (red) methods. 
    Each time step generates $4\times 10^6$ OPs in the three simulations. 
    The computation time of the plasma state and opacity is also shown in the figure with ultramarine color. 
    Note that the EBT calculation is not activated until 25 days after the explosion, and the green curve before day 25 is overlapped to the black curve. }
    \label{fig:timeCompare}
\end{figure}

Figure \ref{fig:timeCompare} shows the computation time of particle propagation at each time step as a function of the time after explosion using the EBT and IBT methods as described in Section \ref{sec:crosscompare}, and the corresponding DCT computation time with $4\times 10^6$ OPs per time step, the same number of OPs as the EBT and IBT results. 
The time needed to compute the plasma state and opacity does not change drastically at different time steps. 
In contrast, the computation time for packet propagation increases in the first $\sim$ 3 days, because the optical depth is high and the OPs cannot easily escape the SN ejecta and undergo repeated interaction events. 
With the expansion of the SN ejecta, the optical depth decreases, and the accumulated OPs escape the SN ejecta, therefore the computation time decreases. 
In particular, the computation time for EBT is a factor of $\sim$ 30 higher than DCT around 25 days after the explosion, and is a factor of $\sim$ 10 higher than DCT around 60 days. 
In contrast, the introduction of IBT only increases the computation time by a factor of 0.3-0.5 of the DCT computation time, throughout the time after the SN explosion, to reach a comparable S/N as EBT. 

\subsection{Signal to Noise Ratio}

In an MCRT simulation, the output spectral flux error scales roughly as $\sigma\propto \sqrt{N}$, where $N$ is the number of photon packets. 
We therefore use the simulation results on the D1D model to measure the simulation error of the three spectral retrieval methods. 
First, we split the output spectra into ultraviolet wavelength range (1000-2000 $\AA$), optical wavelength range (2000-10000 $\AA$), and infrared wavelength range (10000-25000 $\AA$). 
Second, we measure the simulation root mean squared error (RMSE) of the Stokes parameters $\mathcal{Q}$ and $\mathcal{U}$ using the following equation: 

\begin{equation}
    RMSE=\sqrt{\sum_{N_\nu}\frac{ \left(\frac{\mathcal{Q}(\nu)}{\mathcal{I}(\nu)}-0\right)^2 +\left(\frac{\mathcal{U}(\nu)}{\mathcal{I}(\nu)}-0\right)^2}{N_\nu}}\ , 
\end{equation}
where $N_\nu$ is the number of frequency indicies in the selected wavelength range, the summation is over all frequency indicies in the selected wavelength range. 
Note that the theoretical solution of the D1D model is no polarization which results in the zero term in the equation. 
The RMSE observes the inverse square root relation with the number of packets, $RMSE\propto N^{-0.5}$.

\begin{figure*}
    \includegraphics[width=0.33\textwidth]{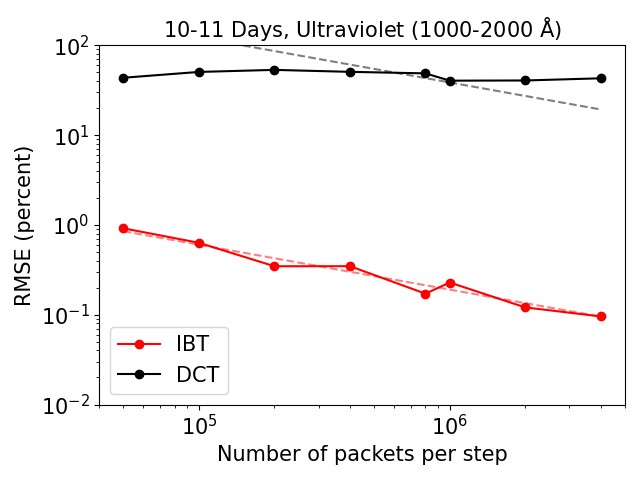}
    \includegraphics[width=0.33\textwidth]{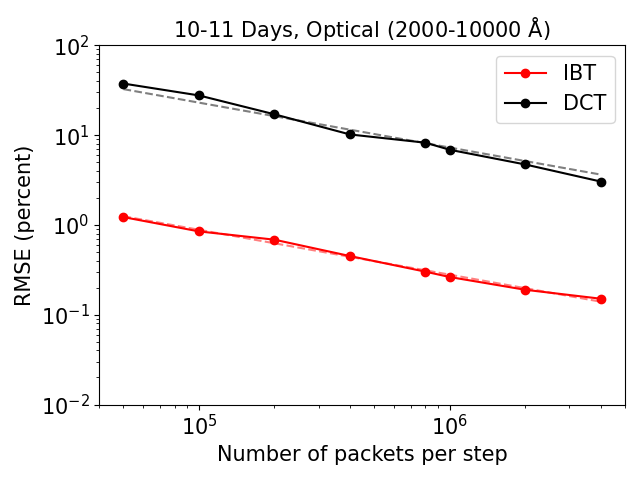}
    \includegraphics[width=0.33\textwidth]{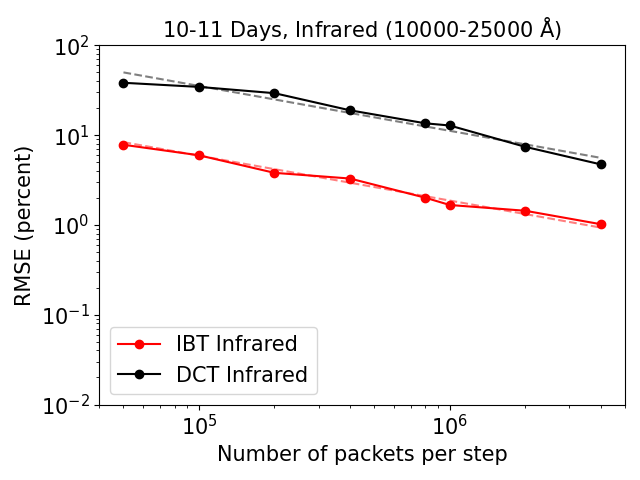}
    \includegraphics[width=0.33\textwidth]{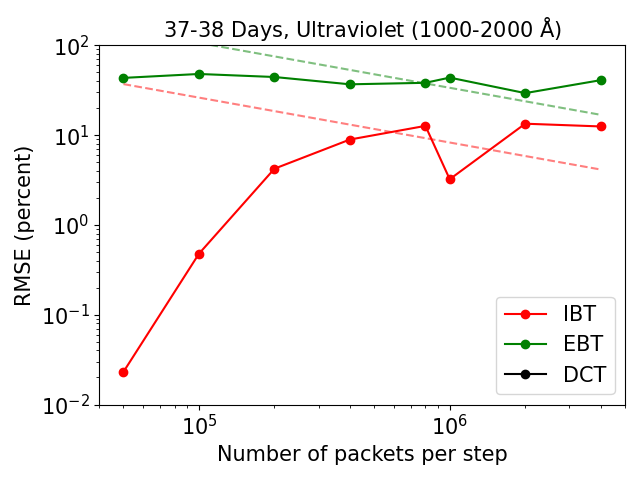}
    \includegraphics[width=0.33\textwidth]{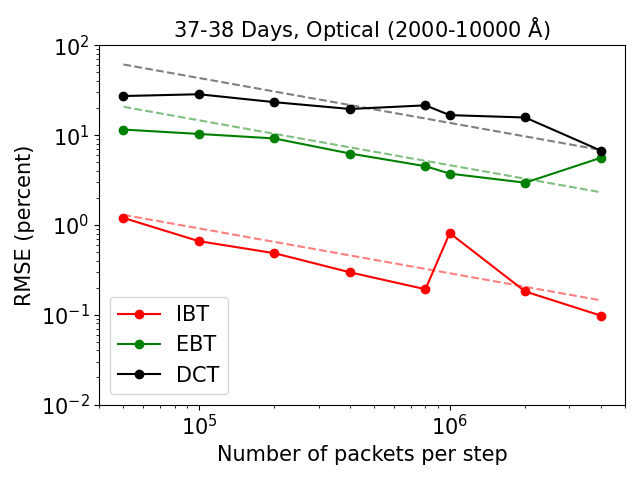}
    \includegraphics[width=0.33\textwidth]{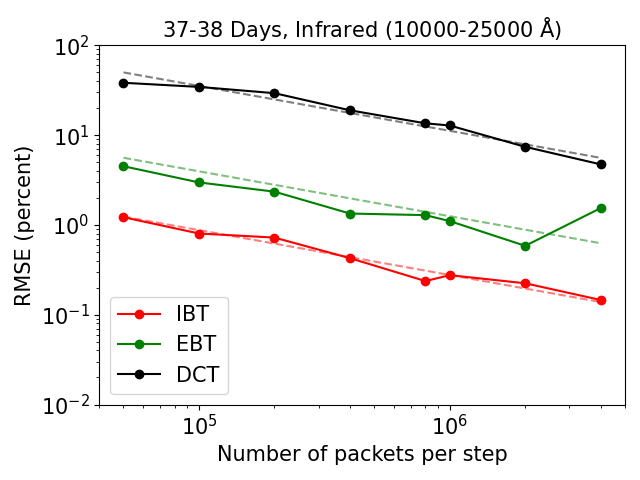}
    \caption{The RMSE of DCT, EBT, and IBT simulation results on the D1D model, as a function of the number of OPs generated per time step. 
    The polarization spectra are binned with 7 pixels. 
    Upper panel shows the RMSE in the time bin of 10-11 days after the explosion, lower panel shows the RMSE in the time bin of 37-38 days after the explosion. 
    From left to right, the wavelength range is ultraviolet, optical, and infrared. 
    The dashed line shows the theoretical inverse square root relation between RMSE and the number of packets ($RMSE\propto N^{-0.5}$) for comparison. }
    \label{fig:rmse}
\end{figure*}

Figure \ref{fig:rmse} shows the relationship between the number of OPs per time step and the resulting RMSE at 10-11 days after the explosion using DCT or IBT, and at 37-38 days after the explosion using DCT, EBT, or IBT. 
We notice the result from IBT could reproduce the $RMSE\propto N^{-0.5}$ relation in most of the time and wavelength ranges. 
The DCT result also reproduces the relation $RMSE\propto N^{-0.5}$ in the optical and infrared wavelength ranges when the number of packets is large enough. 
However, in the ultraviolet wavelength, and in the optical wavelength with the number of packets per step smaller than $\sim 4\times 10^{5}$, the inverse square root relation is not observed due to the large simulation error. 
The result from EBT also reproduces the $RMSE\propto N^{-0.5}$ relation in optical ind infrared wavelength ranges at 37-38 days after the explosion. 
In the ultraviolet wavelength range at 37-38 days after the explosion, none of the methods could simulate good S/N spectra to reproduce the $RMSE\propto N^{-0.5}$ relation. 

A direct comparison of the computation efficiency of the three methods could be made using the RMSE values measured in the optical and infrared wavelength range. 
The RMSE of IBT is smaller than that of DCT by a factor of $20-30$, leading to a factor of $400-900$ less packets needed to calculate to reach similar S/N. 
The RMSE of EBT is smaller than that of DCT by a factor of $5-10$, resulting in a factor of $25-100$ less packets to calculate to achieve a comparable S/N. 
The accuracy comparison between EBT and IBT is consistent with the reports in Table 1 of \cite{Bulla2015VPack}. 

%\section{Application on Observation}\label{sec:05df}

%We apply the IBT method on an observed supernova SN~2005df using the 

%\begin{figure}
%    \centering
%    \includegraphics[width=0.5\textwidth]{fitObs05df.png}
%    \caption{The supernova }
%    \label{fig:05df}
%\end{figure}

%Gesa: This section is intended to add a real supernova observation data versus a simulated spectrum from one of my toy models. SN2005df was a good one but according to some issue, I am not allowed to show this data in the paper. I am waiting for a new supernova from Dr. Yi Yang. 

\section{Conclusion}\label{sec:conclusion}

We present IBT, an algorithm to efficiently synthesize the spectropolarimetry signal from 3-D MCRT simulations. 
Comparing to EBT \citep{Bulla2015VPack}, the present IBT method is upgraded in the following aspects: 

\begin{itemize}
    \item In EBT, a VP is created at each scattering event of an OP, which could lead to an uncontrolled number of VPs and limit the application of EBT on high-opacity plasma models (i.e. SNe Ia at $\sim$3 days after the explosion). However, IBT uses the cell radiance $\mathbf{R}(\Vec{x},t,\nu)$ as a proxy of all VPs in a volume cell of the plasma model, and limits the number of IPs to save memory and computation time in tracing the particles. 
    \item The frequency grid in IBT is logarithmically spaced (Equation \ref{eq:freqGrid}), which simplifies the Doppler shift of cell radiance $\mathbf{R}(\Vec{x},t,\nu)$ to re-indexing the pixel. Therefore, the computation of Equation \ref{eq:updateRad} is simplified to vector operations. This upgrade further accelerates the computation of IP propagation. 
\end{itemize}
%DK2: I would think another important difference in IBT is that each IP actually propagates an entire spectrum, rather than a packet at a single frequency, so this probably helps signal to noise. 
%DK2: One other idea that maybe would make IBT more efficient -- it seems like you emit an IP from every zone at every timesteps, but in some zones the emission will be very low (for example, in very low density corners of the 3D domain. So there really is no point in emitting IPs there, since the emission there wouldn't contribute much at all to the spectrum.  So if you preferrential emitted from zones with more emissivity you probably make things go faster. 
%Gesa2: This feature is actually included in the current code. When the plasma temperature of a cell is 1000K or below, then the cell will not emit an IP. 

In the tests on the D3D model, both IBT and EBT have successfully reproduced the spectral time sequence with spectropolarimetry from 1000\AA\ to 25000\AA , and the S/N is comparable to DCT with $\sim$115 more Monte-Carlo OPs. 
The computation time for IBT is $\sim$ 0.3 of DCT with the same number of OPs per viewing direction, while the computation time for EBT is a factor of $10-30$ of DCT per viewing direction. 
With an opacity limit and wavelength limit to accelerate the computation, the EBT computation time is still about a factor of two of DCT per viewing direction\citep{Bulla2015VPack}. 
Moreover, IBT has successfully calculated the spectral time sequence from 1 day to 60 days after the explosion, while EBT failed before day 25 due to the limited memory space of the computing facility. 

In the tests on the D1D model, IBT has successfully reproduced the zero polarization signal as the theoretical predictions from the spherical symmetry. 
Using the same number of Monte-Carlo OPs, the S/N of the IBT spectrum is $\sim$ 30 times higher than the DCT spectrum S/N and $5-10$ times higher than the EBT spectrum. 

%Several deep-learning based methods are proposed to estimate the SN ejecta structures using radiative transfer programs as aid (e.g. \cite{Chen2020AIAI,Kerzendorf2021Dalek}). 
\cite{Chen2020AIAI,Chen2024ApJ...962..125C} developed the
Artificial Intelligence Assisted Inversion (AIAI) method which enables theoretical models of SNe~Ia to be derived using the observational data as input constraints. 
The published AIAI is based on 1-D models. 
The studies demonstrate that it is possible to combine empirical models of SNe~Ia with detailed radiative transfer code to improve the use of SNe~Ia as cosmological probes.
In the future, we will extend the 1-D models to 3-D time-dependent models. 
The proposed IBT is the ideal algorithm to efficiently generate a time sequence of spectrophotometric and spectropolarimetric data bases which can be used to train AI models, this can be an important step in assimilating the extensive observational data of SNe~Ia into studies of the physics of thermonuclear explosions and the use of SNe~Ia as cosmological distance candles. 

\begin{acknowledgments}

X.C. is supported by the grant NSF-AST 1817099. 
The authors would like to thank Prof. J. C. Wheeler from University of Texas at Austin and Prof. David Jeffery from University of Nevada at Las Vegas for supportive discussions. 
Portions of this research were conducted with the advanced computing resources provided by Texas A\&M High Performance Research Computing. 
This work used FASTER super computer at TAMU HPRC through allocation PHY240215 from the Advanced Cyberinfrastructure Coordination Ecosystem: Services \& Support (ACCESS) program, which is supported by National Science Foundation grants 2138259, 2138286, 2138307, 2137603, and 2138296\citep{boerner2023access}. 

\end{acknowledgments}

%\vspace{5mm}
%\facilities{HST(STIS), Swift(XRT and UVOT), AAVSO, CTIO:1.3m, CTIO:1.5m,CXO}

\software{SEDONA\citep{Kasen2006Sedona}}

\appendix

\section{Rayleigh Scattering}\label{app:scatmat}

The Thomson scattering between free electron and photon is calculated by the Rayleigh scattering phase matrix: 

\begin{equation}
    \boldsymbol{P}_{R}(\Theta)=\frac{3}{4}\begin{pmatrix}
        cos^2\Theta+1 & cos^2\Theta-1 & 0 & 0 \\
        cos^2\Theta-1 & cos^2\Theta+1 & 0 & 0 \\ 
        0 & 0 & 2cos\Theta & 0 \\
        0 & 0 & 0 & 2cos\Theta \\
    \end{pmatrix}\ , 
\end{equation}

where $\Theta$ is the angle between the incident light and the emergent light. 
Before the calculation of Rayleigh scattering phase matrix, the reference axis of the incident light should be rotated into the scattering plane, and the polarization state is changed by the rotation matrix: 

\begin{equation}
    \boldsymbol{L}(\phi_{in})=\begin{pmatrix}
        1 & 0 & 0 & 0 \\
        0 & cos2\phi_{in} & sin2\phi_{in} & 0 \\ 
        0 & -sin2\phi_{in} & cos2\phi_{in} & 0 \\
        0 & 0 & 0 & 1 \\
    \end{pmatrix}\ , 
\end{equation}

where $\phi_{in}$ is the rotation angle between the observer's reference axis and the scattering plane reference axis. 
A similar rotation matrix of $\boldsymbol{L}(\pi-\phi_{out})$ is also applied to the emergent light, in order to convert the polarization state to the observer's reference axis. 

We define $\Bar{\boldsymbol{S}}$ as the multiplication of the packet energy and the dimensionless Stokes vector: 

\begin{equation}
    \Bar{\boldsymbol{S}}=\begin{pmatrix}\Bar{I} \\ \Bar{Q} \\ \Bar{U} \\ \Bar{V}\end{pmatrix}\ = \Bar{E}\Bar{\boldsymbol{s}}\ . 
\end{equation}

Therefore, the full Rayleigh scattering formula in the SN radiative transfer simulation is written as: 

\begin{equation}
    \Bar{\boldsymbol{S}}_{out} = \boldsymbol{L}(\pi-\Bar{\phi}_{out}) \boldsymbol{P}_{R}(\Bar{\Theta}) \boldsymbol{L}(\Bar{\phi}_{in}) \Bar{\boldsymbol{S}}_{in}\ . 
\end{equation}

Note that the Rayleigh scattering happens in the co-moving frame, and all the variables in the above equation are measured in the co-moving frame. 
In Equation \ref{eq:EBTThomScat}, Equation \ref{eq:RadEng}, and Equation \ref{eq:IBTThomScat}, the operator is defined as $\boldsymbol{P}(\Bar{\theta}_{in},\Bar{\phi}_{in},\Bar{\theta}_{out},\Bar{\phi}_{out})=\boldsymbol{L}(\pi-\Bar{\phi}_{out}) \boldsymbol{P}_{R}(\Bar{\Theta}) \boldsymbol{L}(\Bar{\phi}_{in})$. 

Replacing the axis rotation angles and scatter angle ($\Bar{\phi}_{in}$, $\Bar{\phi}_{out}$, $\Bar{\Theta}$) with the propagating direction of the incident OP ($\Bar{\theta}_{in}$, $\Bar{\varphi}_{in}$) and the propagating direction of the emergent VP/IP ($\Bar{\theta}$, $\Bar{\varphi}$), the expression of the operator $\boldsymbol{P}(\Bar{\theta}_{in},\Bar{\phi}_{in},\Bar{\theta},\Bar{\phi})$ becomes \citep{Chandra1960RT}: 

\begin{eqnarray}
    %\Bar{I}_{VP}=1.5*(ll*ll*\Bar{I}+rl*rl*\Bar{Q}+ll*rl*\Bar{U}+lr*lr*\Bar{I}+rr*rr*\Bar{Q}+lr*rr*\Bar{U})\\
    %\Bar{Q}_{VP}=1.5*(ll*ll*I_in[0]+rl*rl*I_in[1]+ll*rl*I_in[2]-lr*lr*I_in[0]-rr*rr*I_in[1]-lr*rr*I_in[2])\\
    %\Bar{U}_{VP}=1.5*(2.0*ll*lr*I_in[0] + 2.0*rr*rl*I_in[1] + (ll*rr + rl*lr)*I_in[2])\\
    %\Bar{V}_{VP}=1.5*(ll*rr - rl*lr)*\Bar{V}
    \Bar{I}_{out}&=&\frac{3}{8\pi}\left(\left((l,l)^2+(l,r)^2\right)\Bar{I}_{in}+\left((r,l)^2+(r,r)^2\right)\Bar{Q}_{in}+\left((l,l)(r,l)+(l,r)(r,r)\right)\Bar{U}_{in}\right)\\
    \Bar{Q}_{out}&=&\frac{3}{8\pi}\left(\left((l,l)^2-(l,r)^2\right)\Bar{I}_{in}+\left((r,l)^2-(r,r)^2\right)\Bar{Q}_{in}+\left((l,l)(r,l)-(l,r)(r,r)\right)\Bar{U}_{in}\right)\\
    \Bar{U}_{out}&=&\frac{3}{8\pi}\left( 2(l,l)(l,r)\Bar{I}_{in}+2(r,r)(r,l)\Bar{Q}_{in}+\left((l,l)(r,r)+(r,l)(l,r)\right)\Bar{U}_{in} \right) \\
    \Bar{V}_{out}&=&\frac{3}{8\pi}\left((l,l)(r,r)-(r,l)(l,r)\right)\Bar{V}_{in}\ , 
\end{eqnarray}

where $(l,l)$, $(r,l)$, $(l,r)$, $(r,r)$ are: 

\begin{eqnarray}
    (l,l)&=&sin(\Bar{\theta})sin(\Bar{\theta_{in}}) + cos(\Bar{\theta})cos(\Bar{\theta_{in}})cos(\Bar{\varphi_{in}}-\Bar{\varphi})\\
    (r,l)&=&cos(\Bar{\theta})sin(\Bar{\varphi_{in}}-\Bar{\varphi})\\
    (l,r)&=&-cos(\Bar{\theta_{in}})sin(\Bar{\varphi_{in}}-\Bar{\varphi})\\
    (r,r)&=&cos(\Bar{\varphi_{in}}-\Bar{\varphi})\ .
\end{eqnarray}

%\bibliography{thebibfile}{}
%\bibliographystyle{aasjournal}

%% This command is needed to show the entire author+affiliation list when
%% the collaboration and author truncation commands are used.  It has to
%% go at the end of the manuscript.
%\allauthors

%% Include this line if you are using the \added, \replaced, \deleted
%% commands to see a summary list of all changes at the end of the article.
%\listofchanges

\end{document}